\documentclass[pre,twocolumn,aps,superscriptaddress,showpacs,floatfix]{revtex4}
\usepackage{graphicx}
\usepackage{dcolumn}
\usepackage{bm}
\usepackage{amsmath}
\usepackage{amssymb}
\begin{document}
\title{Brownian transport in corrugated channels with inertia}
\author{P. K. Ghosh}
\affiliation{Advanced Science Institute, RIKEN, Wako-shi, Saitama,
351-0198, Japan}
\affiliation{Institut f{\"u}r Physik Universit{\"a}t Augsburg,
D-86135 Augsburg, Germany}
\author{P. H\"anggi}
\affiliation{Institut f{\"u}r Physik Universit{\"a}t Augsburg,
D-86135 Augsburg, Germany}
\author{F. Marchesoni}
\affiliation{Advanced Science Institute, RIKEN, Wako-shi, Saitama,
351-0198, Japan}
\affiliation{Dipartimento di Fisica, Universit\`a di Camerino,
I-62032 Camerino, Italy}
\author{F. Nori}
\affiliation{Advanced Science Institute, RIKEN, Wako-shi, Saitama,
351-0198, Japan}
\affiliation{Physics Department, University of Michigan, Ann
Arbor, MI 48109, USA}
\author{G. Schmid}
\affiliation{Institut f{\"u}r Physik Universit{\"a}t Augsburg,
D-86135 Augsburg, Germany}

\begin{abstract}
The transport of suspended Brownian particles dc-driven along
corrugated narrow channels is numerically investigated in the
regime of finite damping. We show that inertial corrections cannot
be neglected as long as the width of the channel bottlenecks is
smaller than an appropriate particle diffusion length, which
depends on the the channel corrugation and the drive intensity.
Being such a diffusion length inversely proportional to the
damping constant, transport through sufficiently narrow
obstructions turns out to be always sensitive to the viscosity of
the suspension fluid. The inertia corrections to the transport
quantifiers, mobility and diffusivity, markedly differ for
smoothly and sharply corrugated channels.

\end{abstract}

\pacs{05.40.-a,05.60.Cd,51.20.+d} \maketitle

\section{Introduction}
\label{intro} 
\begin{figure}[tp]
\centering
\includegraphics[width=0.5\textwidth]{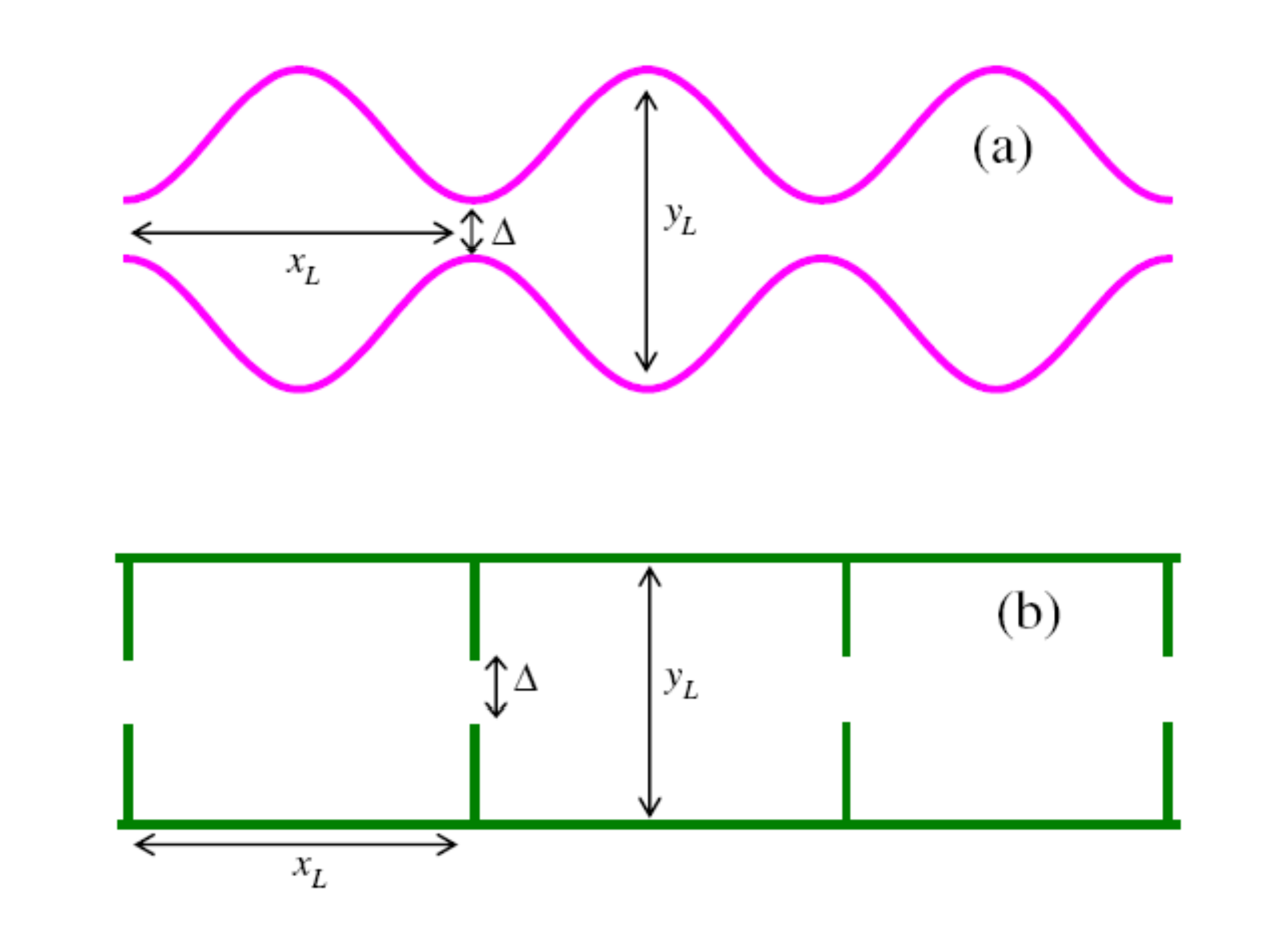}
\caption{(Color online) Sketch of a smoothly corrugated (a)  and a
compartmentalized (b) 2D channel  directed along the $x$-axis. In
both cases the channel unit cell is $x_L$ long and $y_L$ wide; the
radius of the connecting bottlenecks or pores is $\Delta$.
\label{F1}}
\end{figure}

Brownian transport in narrow corrugated channels is
a topic of potential applications to both
natural~\cite{Hille,Kaerger,Brenner} and artificial
devices~\cite{BM}. Depending on the amplitude and geometry of the
wall modulation, corrugated channels fall within two distinct
categories [Fig.~\ref{F1}]: (i) smoothly-corrugated channels
(e.g., as shown in Fig.~1(a)). Also called entropic
channels~\cite{chemphyschem}, these quasi-one-dimensional (1D)
channels were first introduced in Ref.~\cite{Zwanzig} and further
investigated in
Refs.~\cite{Reguera:2001,Kalinay,Laachi,Reguera:2006,Burada,Reichelt,India},
as an instance of two (2D) or three dimensional (3D) systems
describable in terms of an effective 1D kinetic equation. These
are typically modeled as periodic channels with axial symmetry and
unit cells delimited by bottlenecks which are assumed to be narrow
with respect to the cell dimensions, i.e., the channel cross
section and the period; (ii) compartmentalized (or septate)
channels~\cite{Lboro,Bere1,Bere2,PHfest,Borromeo,Shape}.
These septate channels are sharply corrugated channels formed by
identical compartments separated by thin dividing walls and
connected by narrow openings (pores) centered around their axis.
At variance with the case of smoothly corrugated channels,
diffusion in compartmentalized channels cannot be reduced to an
effective 1D kinetic process directed along the axis. Accordingly,
driven transport in such strongly constrained geometries exhibits
distinct features, which cannot be reconciled with the known
properties of Brownian motion in quasi-1D systems
\cite{Brenner,Jacobs,Risken}.

Corrugated channels  are often used to model transport of dilute
mixtures of small particles (like biomolecules, colloids or
magnetic vortices) in confined geometries~\cite{BM}. Each particle
is subjected to thermal fluctuations with temperature $T$ and
large viscous damping constant $\gamma$, and a homogeneous
constant force directed locally parallel to the channel axis. Such
a dc drive is applied from the outside by coupling the particle to
an external field (for instance, by attaching a dielectric or
magnetic dipole, or a magnetic flux to the particle), without
inducing drag effects on the suspension fluid. Interparticle and
hydrodynamic interactions are thus ignored for simplicity (for a
discussion on this assumption see Ref.~\cite{chemphyschem}).

\begin{figure*}[tp]
\centering
\includegraphics[width=0.48\textwidth]{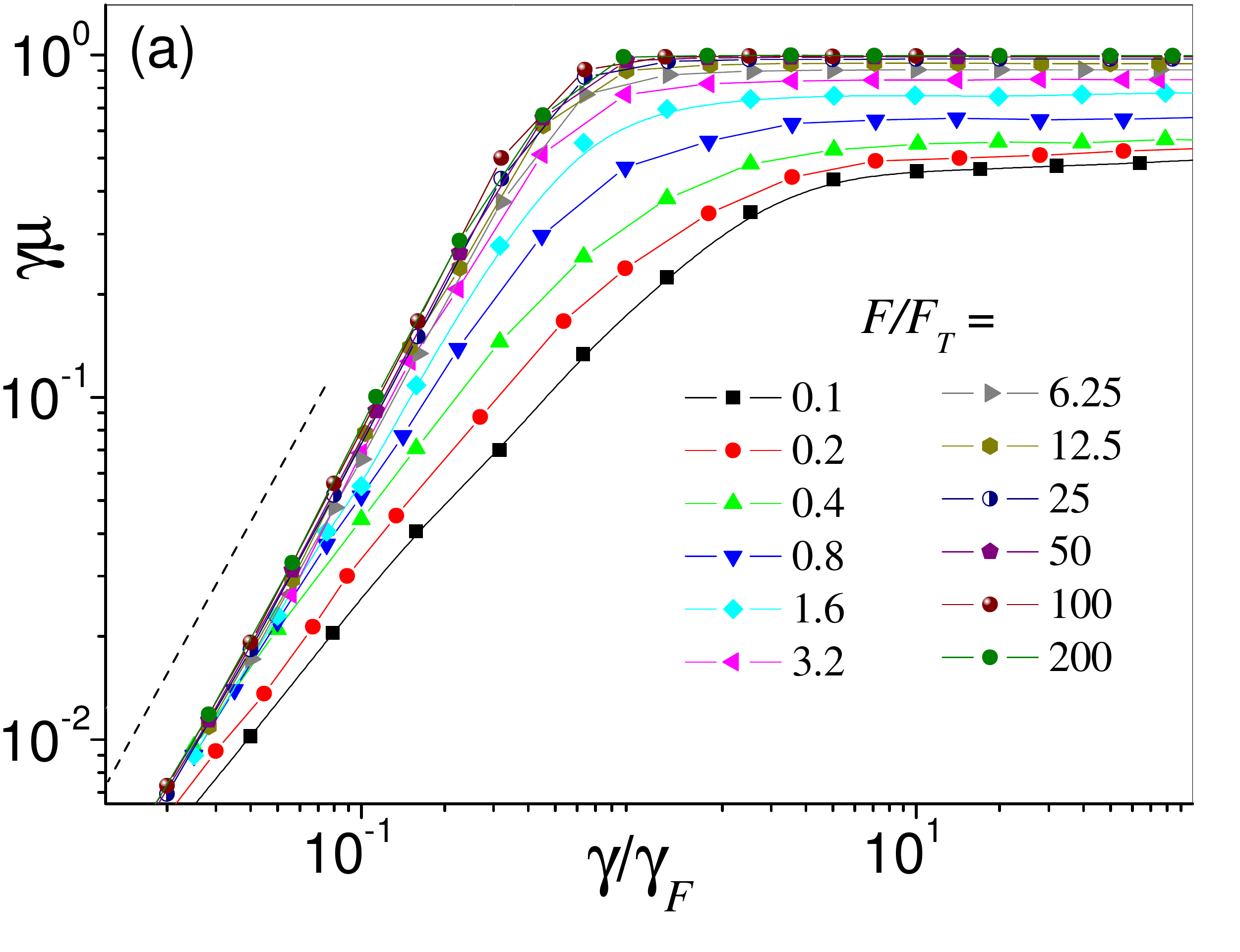}
\includegraphics[width=0.48\textwidth]{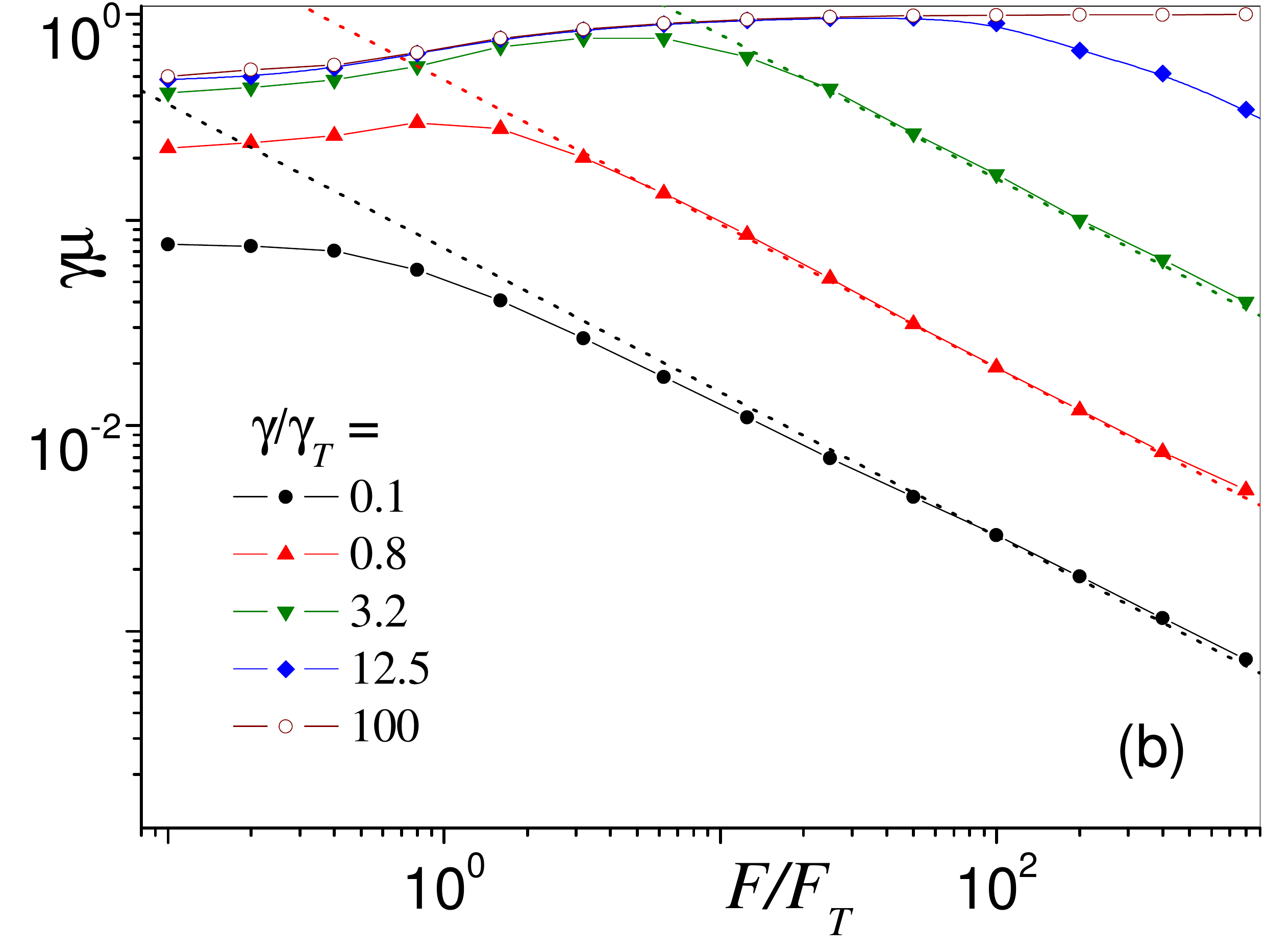}
\includegraphics[width=0.48\textwidth]{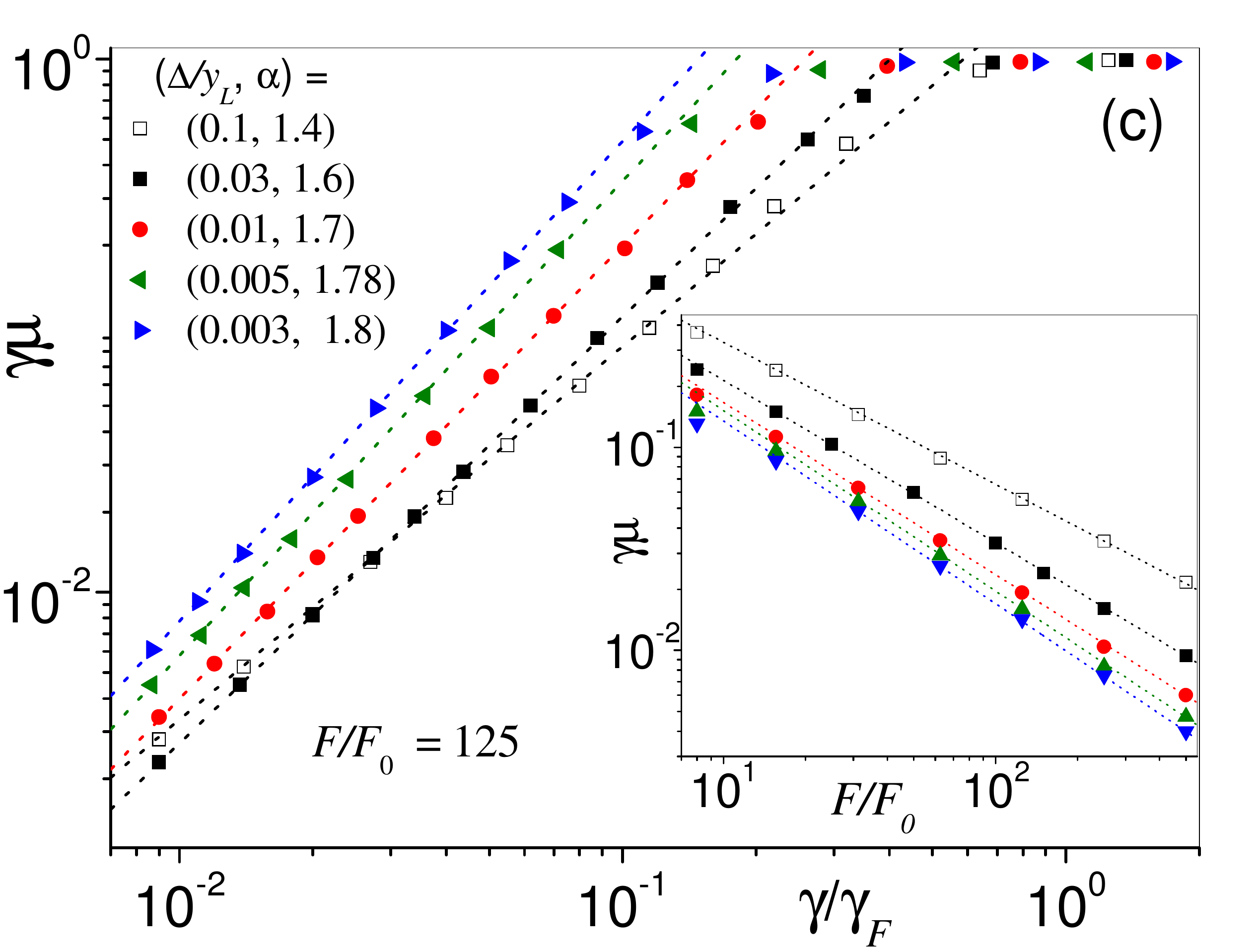}
\includegraphics[width=0.48\textwidth]{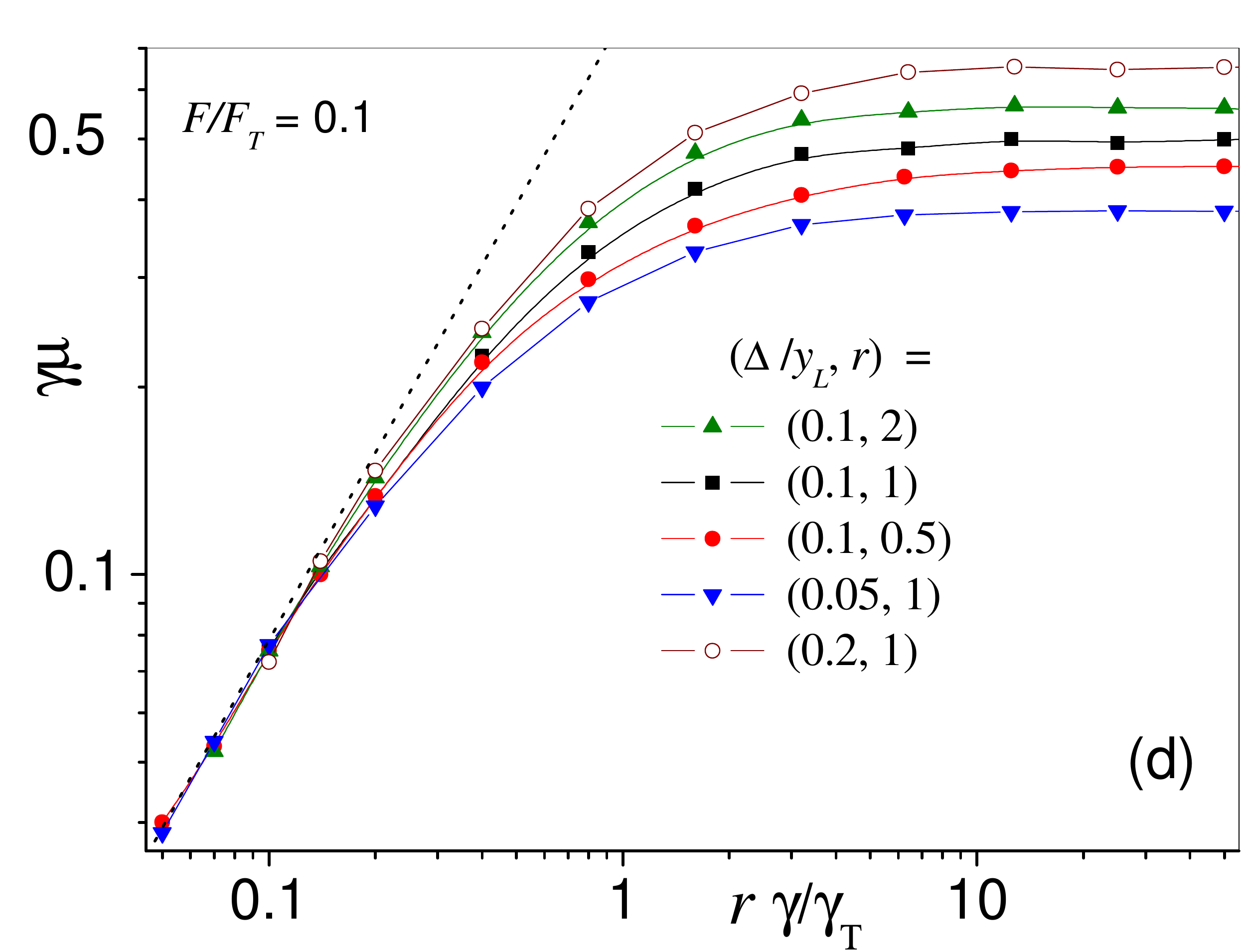}
\caption{(Color online) Rescaled mobility, $\gamma \mu$, in a
smoothly corrugated channel with $r=x_L/y_L=1$, $\Delta/y_L=0.1$,
and (a) versus $\gamma/\gamma_F$ for different $F$; (b) versus
$F/F_T$ for different $\gamma$. The relevant scaling parameters
are $F_T=kT/\Delta$, $\gamma_T=\sqrt{mkT}/\Delta$, Eq.~(\ref{gT}),
and  $\gamma_F=\sqrt{mF/\Delta}$, Eq.~(\ref{gF}). The dashed lines
represent, respectively, the fitting power laws
$(\gamma/\gamma_F)^\alpha$ in (a) and  $(F/F_T)^{-\alpha/2}$ in
(b), both with $\alpha=1.4$ (see Sec.~\ref{corrugated}). In (c)
$\gamma \mu$ is plotted versus $\gamma/\gamma_F$ (main panel) and
$F/F_0$ (inset) for $\gamma/\gamma_T=0.8$, $F/F_T=125$, and
different cross-section ratios, $\Delta/y_L$. The corresponding
fitting exponents $\alpha$ are also reported in the legend. In the
inset, $F$ is expressed in units of $F_0$ instead of $F_T$ for
graphical reasons. The dependence of $\gamma \mu$ on the geometry
of the channel unit cell for low damping and small drives is
illustrated in (d), where $\gamma \mu$ is plotted versus
$r\gamma/\gamma_T$. The predicted linear law with slope $\pi/4$
\cite{our EPL} is represented by a dotted line [see also
Eq.~(\ref{th-mu0}) and text following]. \label{F2}}
\end{figure*}

In this paper  we investigate the relevance of the inertia effects
due to the viscosity of the suspended particle. As is often the
case with biological and most artificial suspensions~\cite{BM},
the Brownian particle dynamics in the bulk can be regarded as
overdamped. This corresponds to (i) formally setting the mass of
the particle to zero, $m=0$, or, equivalently, to make the
friction strength $\gamma$ tend to infinity, and (ii) assuming $F$
smaller than the thermal force $F_0=\gamma \sqrt{kT/m}$
(Smoluchowski approximation)~\cite{SEQ}. The current literature on
corrugated channels invariably assumes such an overdamped limit.
But how large is an infinite $\gamma$ (or how small a zero $m$)?
The answer, of course, depends on the geometry of the channel.

 Our main conclusion is that the overdamped dynamics assumption for
Brownian diffusion through pores of width $\Delta$ subjected to a
homogeneous drive $F$, applies only for the conditions~\cite{SEQ}:
$$\gamma \gg
\sqrt{mkT}/\Delta,\;\;\;\; {\rm and} \;\;\;\;  \gamma \gg
\sqrt{mF/\Delta},$$ irrespective of the degree of corrugation.
This means that inertial correction cannot be neglected as long as
the  Brownian diffusion is spatially correlated on a length
($l_T=\sqrt{mkT}/\gamma$ at small dc drive, or $l_F=mF/\gamma^2$
at large  dc drive) of the order of, or larger than, the pore
width $\Delta$. Therefore, for sufficiently narrow pores or
sufficiently large drives, inertia always comes into play by
enhancing the blocking action of the channel bottlenecks.

This paper is organized as follows.  In Sec.~\ref{model} we
introduce the Langevin equation formalism employed in our
simulation code.  Simulation data for the particle mobility and
diffusivity are analyzed in Sec.~\ref{corrugated} as functions of
the drive, the channel geometry, and the damping constant in
sinusoidally-corrugated channels. We report significant deviations
from the best known overdamped regime. In Sec.~\ref{septate} we
consider the case of septate channels for which dependable fitting
formulas could be analytically obtained. Inertial effects in these
two limiting corrugation regimes are compared in Sec.
\ref{discussion}. Finally, in Sec.~\ref{conclusion} we add some
concluding remarks.

\section{Model} \label{model}

Let us consider a point-like Brownian particle  of mass $m$
diffusing in a 2D suspension fluid contained in a periodic channel
with unit cell $x_L \times y_L$, as illustrated in Fig.~\ref{F1}.
The particle is subjected to a homogeneous force ${\vec F}$. The
damped dynamics of the particle is modeled by the 2D Langevin
equation,
\begin{equation}\label{le}
m\frac{d^2{\vec r}}{dt^2}=-\gamma \frac{d{\vec r}}{dt}+{\vec F}\;+ \sqrt{\gamma kT}~{\vec \xi}(t),
\end{equation}
where ${\vec r}=(x,y)$. The random forces ${\vec
\xi}(t)=(\xi_x(t),\xi_y(t))$ are zero-mean, white Gaussian noises
with autocorrelation functions $\langle \xi_i(t)\xi_j(t')\rangle =
2\delta_{ij}\delta(t-t')$, with $i,j=x,y$. Here, $\gamma$ plays
the role of an effective viscous damping constant incorporating
all additional effects that are not explicitly accounted for in
Eq.~(\ref{le}), like hydrodynamic drag, particle-wall
interactions, etc. We numerically integrated Eq.~(\ref{le}) by a
Milstein algorithm~\cite{Milstein}. The stochastic averages
reported in the forthcoming sections were obtained as ensemble
averages over 10$^6$ trajectories with random initial conditions;
transient effects were estimated and subtracted.

As anticipated in Sec.~\ref{intro}, we considered two categories
of periodic channels, smoothly-corrugated and septate channels.
The symmetric walls of smoothly-corrugated channels have been
modeled by the sinusoidal functions $\pm w(x)$, where
\begin{equation}
w(x)=\frac{1}{4}\left [ (y_L+\Delta)-(y_L-\Delta)\cos\left (\frac{2\pi x}{x_L}\right)\right ],
\label{walls}
\end{equation}
as shown in Fig.~\ref{F1}(a)]. On the other, the compartments of
the septate channels are rectangular and the dividing walls have
zero width, as shown in Fig.~\ref{F1}(b).

Two quantifiers have been used to best represent the different transport properties of these two channel geometries in the overdamped limit, $\gamma \to \infty$:

{\it (i) mobility.} The response of a Brownian particle in a
channel subjected to a dc drive, $F$, oriented along the $x$-axis
is expressed by the mobility,
\begin{equation}
\mu(F) = \langle v(F) \rangle /F,
\label{mobility}
\end{equation}
where $\langle v\rangle \equiv \langle \dot x(F) \rangle
=\lim_{t\to\infty}[\langle x (t) \rangle -x(0)]/t$. In entropic
channels, $\mu(F)$ increases from a relatively small value
$\mu_0$, for $F=0$,   up to the free-particle limit, $\gamma
\mu_\infty=1$, for $F\to \infty$ \cite{Burada}.  We recall that in
a smooth channel a free particle drifts with speed
$v_\infty=F/\gamma$, that is, with $\gamma \mu=1$. On the
contrary, in compartmentalized channels $\gamma \mu(F)$ decreases
monotonically with increasing $F$ towards a geometry-dependent
asymptotic value, $\gamma \mu_\infty$, equal to the ratio of the
pore to the channel cross-section~\cite{Lboro}, that is
\begin{equation}
\gamma\mu_\infty = \Delta/y_L,
\label{mobOO}
\end{equation}

{\it (ii) diffusivity.} As a Brownian particle is driven across a periodic array of bottlenecks or compartment pores, its diffusivity,
\begin{equation}
D(F) = \lim_{t \to \infty}[\langle x^2(t)\rangle -
\langle x(t) \rangle^2]/2t,
\label{diffusivity}
\end{equation}
picks up a distinct $F$-dependence. In entropic channels with
smooth bottlenecks, for $F\to \infty$ the function $D(F)$
approaches the free diffusion limit, $D(\infty)=D_0$, after going
through an excess diffusion peak centered around an intermediate
(temperature dependent~\cite{Burada}) value of the drive. The bulk
or free diffusivity, $D_0$, is proportional to the temperature,
$D_0=kT/m\gamma$. Such a peak signals the depinning of the
particle from the entropic barrier array \cite{Costantini}. In
compartmentalized channels, instead, $D(F)$ exhibits a distinct
quadratic dependence on $F$ \cite{Bere1,PHfest}, reminiscent of
Taylor's diffusion in hydrodynamics~\cite{Taylor}, that is, for
$\Delta \ll y_L$,
\begin{equation}\label{th-diffOO}
\frac{D(F)}{D_0}=\frac{1}{2}\left (\frac{F \Delta}{kT}\right )^2.
\end{equation}
This observation suggests that the particle never frees itself from the geometric constriction of the compartment pores, no matter how strong $F$.

These two quantifiers can also be used to assess the magnitude of
the inertia effects. We remind here that, in the absence of
external drives and for any value of the damping constant,
Einstein's relation~\cite{Risken}
\begin{equation}
\gamma \mu_0 = D(0)/D_0,
\label{einstein}
\end{equation}
establishes the dependence of the transport parameters on the
temperature and the channel compartment geometry under equilibrium
conditions~\cite{Lboro}.
\begin{tiny}
\begin{table*}
\caption {\label{table1} Summary of characteristic scaling
parameters and their meaning}

\begin{tabular}{|l|l|}
  \hline

  $~F_0 \;= \gamma\sqrt{\frac{kT}{m}}$  & \begin{tabular}{l}\\ Thermal force: viscous force experienced by a Brownian particle \\ with thermal velocity
  $v_{\rm th}=\sqrt{kT/m}$.\\ \\
    \end{tabular}  \\
  \hline
  $~D_0 \;= \frac{kT}{\gamma}, ~ v_\infty=\frac{F}{\gamma}$  &  \begin{tabular}{l} \\ Free diffusivity and velocity in bulk.\\ \\ \end{tabular}  \\
  \hline
  $~l_T \;= \;\frac{\sqrt{mkT}}{\gamma}$  & \begin{tabular}{l} \\ Thermal length: distance covered by a Brownian particle diffusing
   \\ with thermal  velocity $v_{\rm th}$ in the relaxation time,
   $m/\gamma$.\\ \\
   \end{tabular}  \\
  \hline
 $~l_F \;= \;\frac{mF}{\gamma^2}$  & \begin{tabular}{l} \\ Ballistic length: distance covered by a driven Brownian particle
  \\ drifting with velocity $v_\infty$ in the relaxation time, $m/\gamma$. \\ \\ \end{tabular}  \\
  \hline
  $~\gamma_T \;= \;\frac{\sqrt{mkT}}{\Delta}$  & \begin{tabular}{l} \\ Damping cut-off at the pore (zero drive): $l_T=\Delta$. \\ \\ \end{tabular}  \\
  \hline
 $~\gamma_F = \sqrt{\frac{m F}{\Delta}}$ & \begin{tabular}{l} \\ Damping cut-off at the pore (strong drive): $l_F=\Delta$.\\ \\
 \end{tabular}  \\
  \hline
$~D_T \;= \frac{kT}{\gamma_T}$, $ F_T = \frac{kT}{\Delta}$  &
\begin{tabular}{l} \\ Scaling parameters introduced in Figs. \ref{F2}-\ref{F5}; \\ obtained by replacing $\gamma$ with $\gamma_T$, respectively,
 in $D_0$ and $F_0$. \\ \\  \end{tabular}  \\
  \hline

\end{tabular}
\end{table*}
\end{tiny}

In preparation for the quantitative analysis of our numerical
data, we remark that Eq.~(\ref{le}) can be conveniently rewritten
in terms of the rescaled units $t \to \gamma t/m$ and $x/l_T$,
with $l_T = \sqrt{mkT}/\gamma$. A straightforward dimensional
argument shows that, for any given channel unit cell $x_L\times
y_L$, both the particle rescaled mobility, $\gamma\mu$,
Eq.~(\ref{mobility}), and its rescaled diffusivity, $D/D_0$,
Eq.~(\ref{diffusivity}), are functions of the rescaled drive,
$F/F_0$, with $F_0=\gamma \sqrt{kT/m}$, and three cell parameters,
typically, the pore width, $\Delta/l_T$, the pore-to-channel
cross-section ratio, $\Delta/y_L$, and the compartment
aspect-ratio, $r=x_L/y_L$ (see Table~\ref{table1}). Note that a
simultaneous rescaling of all lengths by a factor $\kappa$ would
correspond to a noise intensity rescaling, $T \to T/\kappa^2$.
Throughout our simulations we assumed narrow channels with small
bottlenecks, meaning that $x_L\geq y_L$ and $\Delta \ll y_L$.

\begin{figure}[tp]
\centering
\includegraphics[width=0.45\textwidth]{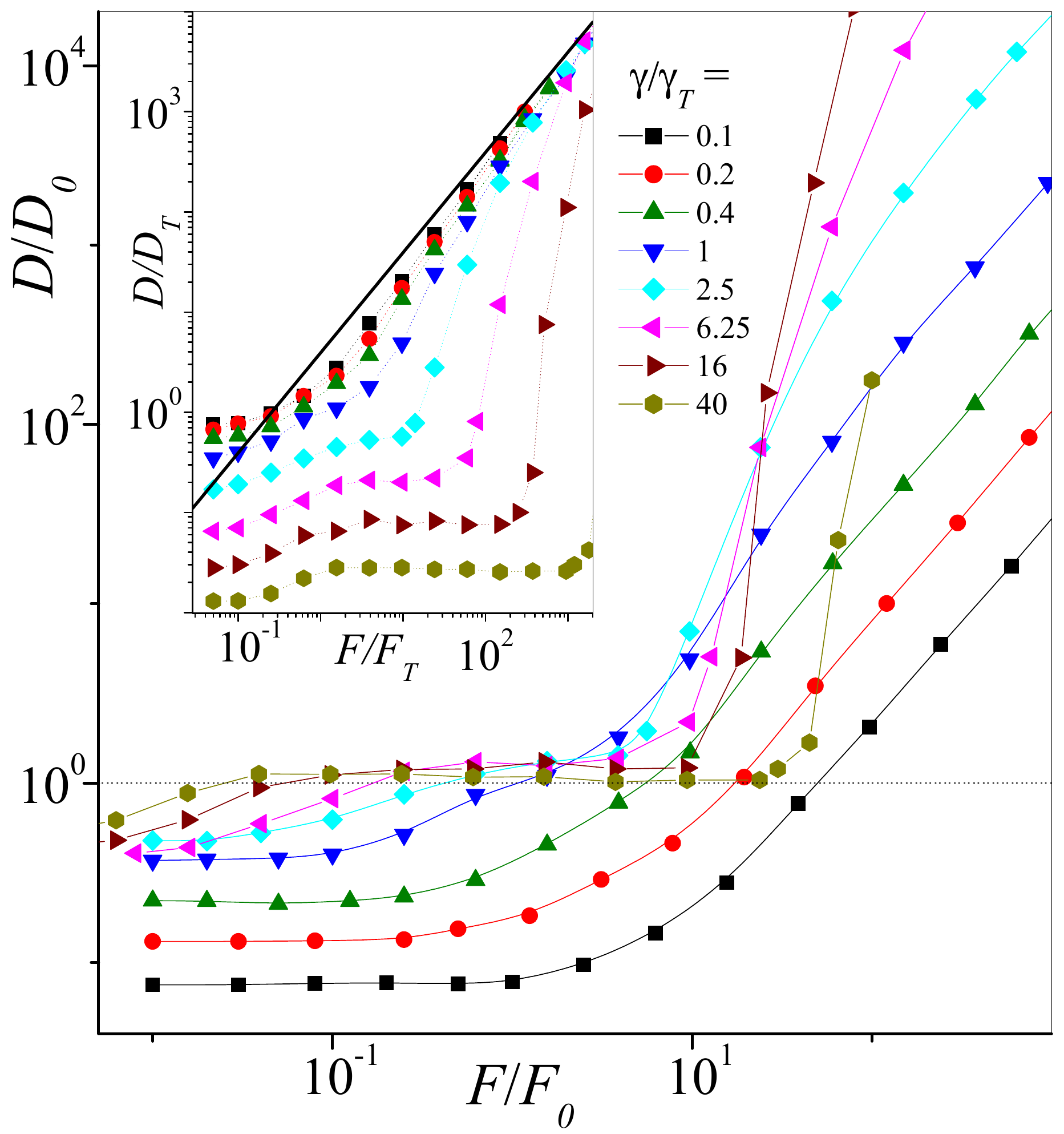}
\caption{(Color online) Rescaled diffusivity, $D/D_0$, {\it
versus} $F/F_T$ (main panel) and $D/D_T$ versus $F/F_0$ (inset) in
the corrugated channel of Eq.~(\ref{walls}) with $r=1$,
$\Delta/y_L=0.1$, and different $\gamma$. The scaling parameters
introduced here are  $D_T= kT/\gamma_T$ and $F_0=\gamma
\sqrt{kT/m}$. The solid line in the inset is the heuristic power
law of Eq.~(\ref{corrugated-diff}). \label{F3}}
\end{figure}

\section{Corrugated channels} \label{corrugated}

As anticipated in a preliminary report \cite{our EPL}, inertial effects  in corrugated channels become apparent both for small $\gamma$ and for large $F$. By inspecting Fig. \ref{F2} we immediately realize that inertia tends to suppress the particle mobility through the channel bottlenecks. Indeed, in the underdamped limit, $\gamma \to 0$, the rescaled mobility drops to zero, no matter what $F$ [Fig. \ref{F2}(a)]. 
In particular, when expressing $\gamma$ in units of
$\gamma_F=\sqrt{mF/\Delta}$, see Eq.~(\ref{gF}) below, the
mobility curves at large drives tend to collapse on a universal
curve well fitted by the power law $(\gamma/\gamma_F)^\alpha $
with $\alpha=1.4$. Correspondingly, in Fig.~\ref{F2}(b) the
mobility  decays like $F^{-\alpha/{2}}$ for small $F\gg F_T$.

The power law, $\gamma \mu \propto (\gamma/\gamma_F)^\alpha$,
introduced here is only a convenient fit of the rescaled mobility
function, even if it holds for two or more decades of
$\gamma/\gamma_F$. [Note that the power law $\gamma \mu \propto
F^{-\alpha/2}$ works throughout the entire $F$ range explored in
Fig.~\ref{F2}(b).] The analytical form of that function remains to
be determined. The data reported in Fig.~\ref{F2}(c) clearly
suggests that the fitting exponent $\alpha$ slightly depends on
$\Delta$, with $\alpha \to 2$ in the limit $\Delta \to 0$.

The dependence of the rescaled mobility on the system parameters in the underdamped limit is further illustrated in Fig. \ref{F2}(d), where at low $\gamma$ and for vanishingly small drives, the mobility grows proportional to the aspect ratio $r=x_L/y_L$ of the channel unit cell and the pore cross section $\Delta$.

Deviations from the expected overdamped behavior are the more prominent in the diffusivity data. As shown in Fig. \ref{F3}, at large $\gamma$ the curves $D(F)$ approach the horizontal asymptote $D(F)=D_0$, as expected \cite{Burada}. However, beyond a certain value of $F$, seemingly proportional to $\gamma^2$ (inset), these curves abruptly part from their horizontal asymptote with a sort of cusp. In the underdamped limit, the $F$ dependence of the diffusivity bears no resemblance with the typical overdamped behavior. At low $\gamma$, all $D(F)$ data sets collapse on a unique curve [Fig. \ref{F3}, inset], which tends to a value smaller than $D_0$ for $F\to 0$, and diverges for $F \to \infty$, like $F^\beta$ with $\beta\simeq 1$. Such power law holds for large $\gamma$, as well, though for sufficiently large $F$, only. Indeed, for exceedingly large $F$, all $D(F)$ curves seem to eventually approach a unique asymptote, irrespective of $\gamma$.

\begin{figure}
\centering
\includegraphics[width=0.44\textwidth]{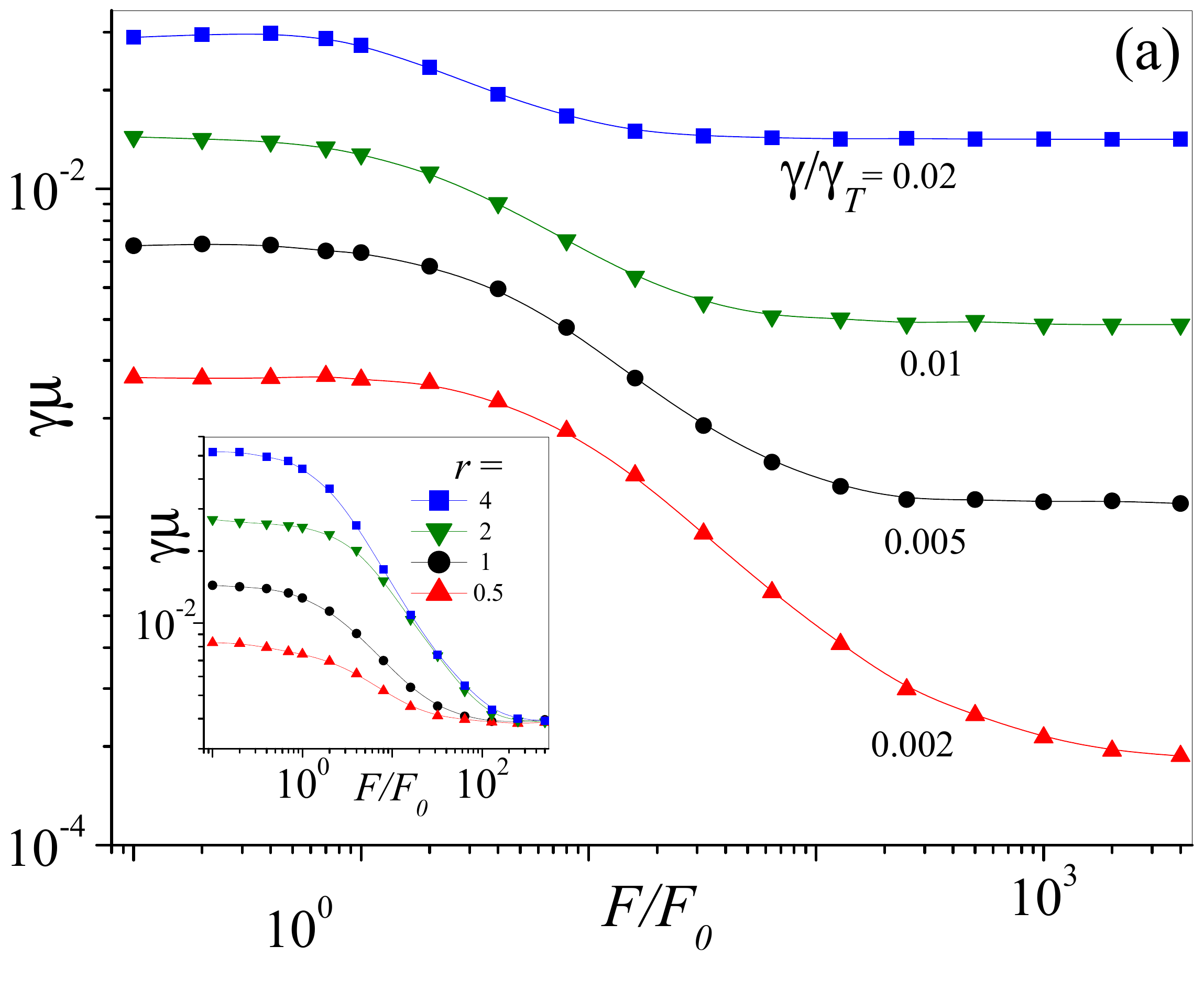}
\includegraphics[width=0.44\textwidth]{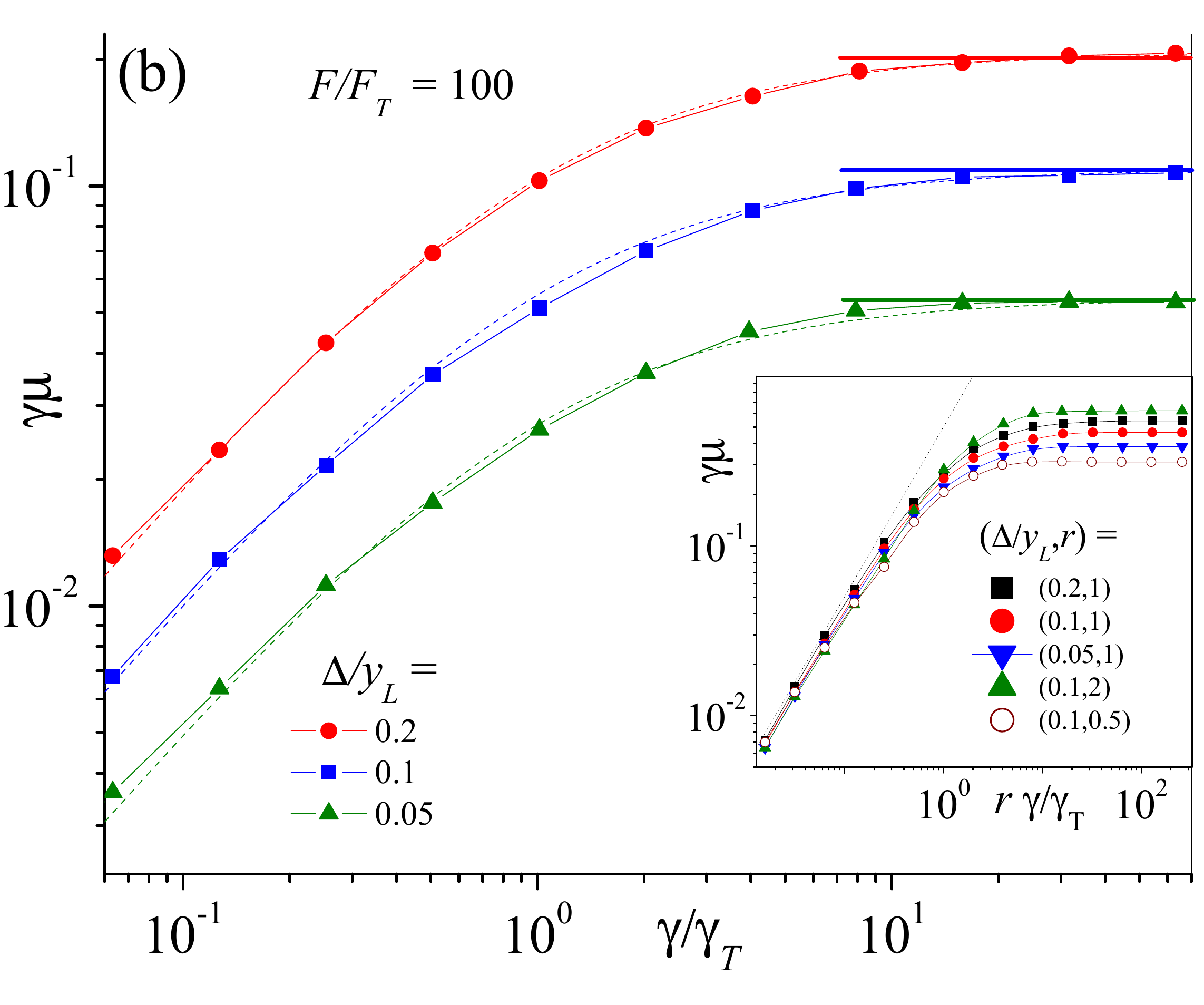}
\includegraphics[width=0.44\textwidth]{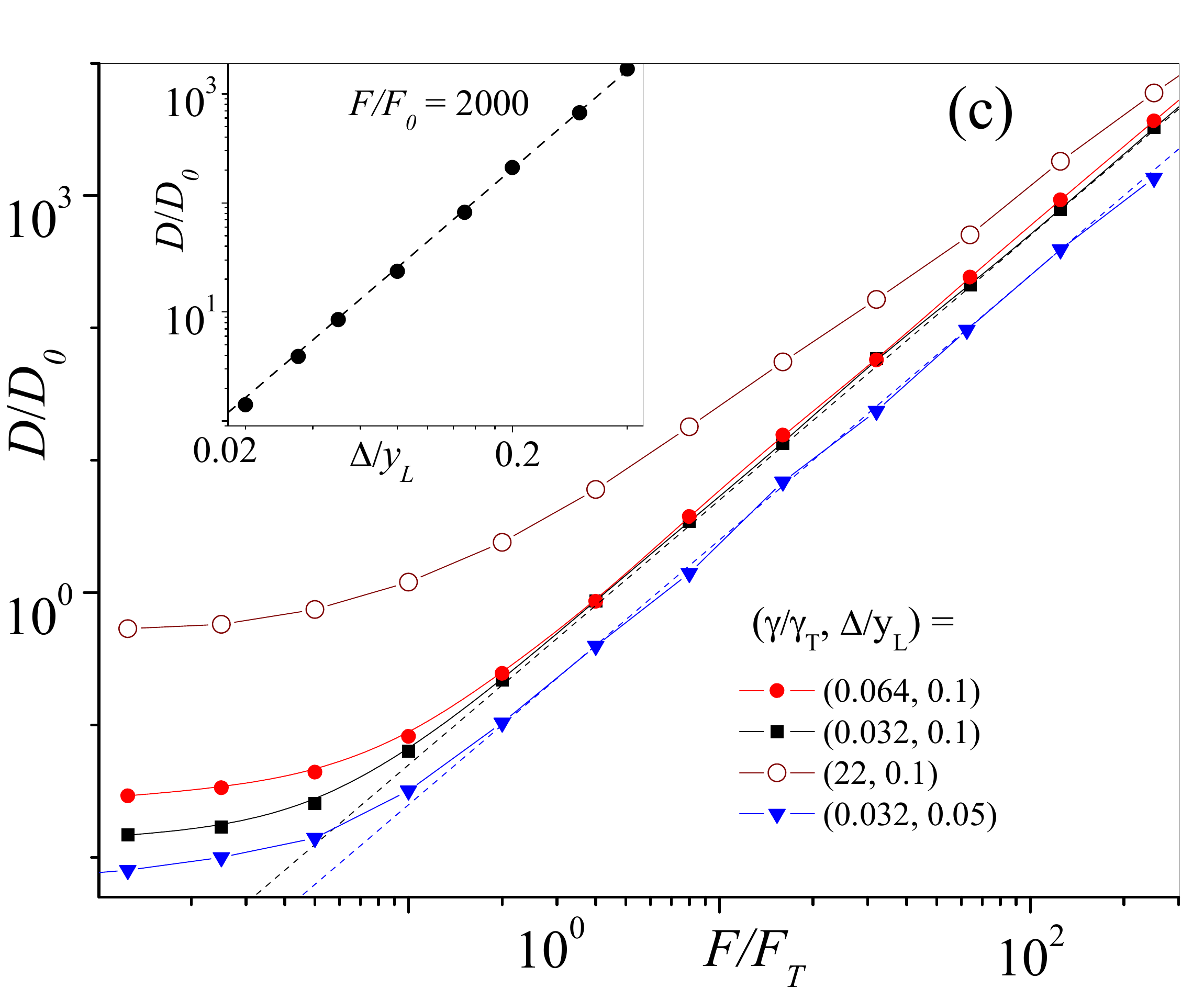}
\caption{(Color online) Transport in a compartmentalized channel
with $r=x_L/y_L=1$: rescaled mobility, $\gamma \mu$, versus
$F/F_0$ (a) and versus $\gamma/\gamma_T$ (b); diffusivity,
$D/D_0$, versus $F/F_T$ (c). The remaining simulation parameters
are reported in the legends. The relevant scaling parameters are
$F_0=\gamma \sqrt{kT/m}$, $F_T=kT/\Delta$, and
$\gamma_T=\sqrt{mkT}/\Delta$. Inset of (a): $\gamma \mu$ versus
$F/F_T$ for different $r$ and $\Delta/y_L$. Inset of (b): $\gamma
\mu$ versus $\gamma/\gamma_T$ for different $\Delta/y_L$ and $r$.
Inset of (c): $D/D_0$ versus $\Delta/y_L$ for $F/F_T=2~10^3$. The
dotted curves represent the approximate analytical expressions of
Eqs.~(\ref{th-mob}) and (\ref{th-diff}), respectively for the
mobility in (b) and the diffusivity in (c) (main panel and
insets). In (b) the quantity $\gamma \mu|_\infty$ was estimated
from the horizontal asymptotes. Note that $\gamma \mu_\infty$ (see
Sec.~\ref{model}) is known to be proportional to $\Delta$ for $F
\to \infty$ [horizontal arrows, Eq.~(\ref{mobOO})], and to $|\ln
\Delta|$ for $F \to 0$ \cite{PHfest}. \label{F4}}
\end{figure}

By comparing the plots of Figs.~\ref{F2}-\ref{F3} we conjecture
that corrections due to inertia become significant in two regimes,
namely:\\
{(i)} at {\it low drives} for
\begin{equation}
\gamma\lesssim \gamma_T=\sqrt{mkT}/\Delta \;.
\label{gT}
\end{equation}
This characteristic damping was used to rescale the mobility data
in Fig.~\ref{F1} [see also Fig.~\ref{F2}(b), inset]; moreover, in
Fig.~\ref{F3}, for $\gamma <\gamma_T$ the diffusivity becomes a
monotonic function of $F$ with no plateau around $D_0$. The
physical meaning of $\gamma_T$ is simple. For $\gamma<\gamma_T$
the {\em thermal} length $l_T=\sqrt{mkT}/\gamma$ grows larger than
the width of the pores, $\Delta$, so that the Brownian particle
cannot reach the normal diffusion regime, implicit in Einstein's
relation, before bouncing off the pore walls. As a consequence,
the Smoluchowski approximation fails in the vicinity of the
bottlenecks.

Replacing $\gamma$ with $\gamma_T$ in the bulk quantities $D_0$
and $F_0$ yields, respectively, $D_T=kT/\gamma_T$ and
$F_T=kT/\Delta$. These are the $\gamma$-independent rescaling
factors introduced in Figs.~\ref{F2}-\ref{F3} to characterize the
inertia effects of the pore constrictions;
\\
{(ii)} at {\it high drives} for
\begin{equation}
\gamma\lesssim \gamma_F=\sqrt{mF/\Delta}\;.
\label{gF}
\end{equation}
As pointed out in Ref.~\cite{PHfest}, the large-drive regime sets
on when the length scale of the longitudinal particle distribution
grows smaller than the pore size, namely for $F \gg F_T$. In the presence a strong dc drive, the condition  $\gamma \gg \gamma_T$ does not suffices to ensure normal diffusion: the additional condition that $\Delta \gg l_F$ is required. Here, $l_F=mF/\gamma^2$ represents the {\em ballistic} length of a driven-damped particle, that is an estimate of the bouncing amplitude of a driven particle against the bottleneck. 
Upon increasing $F$ at constant $\gamma$, $l_F$ eventually grows
larger than $\Delta$ and inertia comes into play. This mechanism
is clearly responsible for the abruptly increasing branches of
$D(F)$ in Fig.~\ref{F3}. A synoptic comparison of all
characteristic scaling parameters of the system is displayed in
Table \ref{table1}.

In conclusion, the low- and large-drive limits are quantitatively
defined as $F\ll F_T$ and $F\gg F_T$, respectively. As $\gamma_F$
was introduced to characterize the large drive (ballistic) regime,
clearly $\gamma_F > \gamma_T$. This means that applying a large
external drive makes the effects of inertia stronger. On the other
hand, if we decrease $\Delta$, while keeping $F$ constant, inertia
effects are controlled by $\gamma_T$ rather than by $\gamma_F$, as
eventually $\gamma_T > \gamma_F$. The smooth crossover between
these two regimes is responsible for the weak $\Delta$ dependence
of the fitting exponent $\alpha$ in Fig.~\ref{F2}.

An analytical derivation of the transport quantifiers in the presence of strong inertial effects (low $\gamma$ and/or large $F$) proved a difficult task. This is the case, for instance, of the universal mobility curve in the inset of Fig.~\ref{F2}(a). 
To gain a deeper insight on this and related issues we address next the particular case of a rectangular compartmentalized channel.

\section{Septate channels} \label{septate}

The role of inertia in compartmentalized channels is illustrated
by the plots of Fig.~\ref{F4}. In panel (a) the rescaled mobility
curve $\gamma \mu(F)$ at low damping exhibits a horizontal
asymptote for $F\to \infty$. However, in comparison with the
overdamped case reported in Sec.~\ref{model}, such an asymptote is
proportional to $\Delta$ only for relatively narrow pores (see
also inset) and is strongly suppressed with decreasing $\gamma$.
The dependence of the mobility on the damping constant is better
illustrated in panel (b), where $\gamma \mu$ linearly increases
with $\gamma$ before reaching the limit predicted in the
overdamped regime \cite{Lboro,Bere1,Bere2,PHfest}. Similar
behaviors were observed both at low (inset) and large drives (main
panel). For large drives the rescaled mobility actually converges
toward the estimate $\gamma \mu|_\infty$ of Eq.~(\ref{mobOO}).

The dependence of the rescaled mobility on the compartment
geometry is further illustrated in the inset of Fig.~\ref{F4}(b):
in the zero-drive limit and for low $\gamma$, the mobility is
proportional to the aspect ratio $r=x_L/y_L$ and the pore size,
$\Delta$, as already reported for the corrugated channels of
Fig.\ref{F2}(d).

Contrary to the smoothly corrugated channels of
Sec.~\ref{corrugated}, the drive dependence of the diffusivity is
apparently not much affected by inertia. As shown in
Fig.~\ref{F4}(c), the curves $D(F)$ keep diverging quadratically
with $F$, irrespective of the compartment size and the damping
constant, like in the overdamped limit. In the notation introduced
above for the corrugated channel, $D(F)$ scales like $F^\beta$
but, contrary to Fig.~\ref{F3}, here $\beta=2$.  The power-law
dependence of $D(F)$ on the pore size and the channel width is
displayed in the insets of Fig.~\ref{F4}(c).

Although of lesser applicability, septate channels have a
practical advantage over smoothly corrugated channels, as they are
characterized by distinct time scales, which often allow
convenient analytical approximations. The problem under study is
no exception.

Let us consider first the rescaled mobility at large drives, $F\gg
F_T$. As anticipated in Secs.~\ref{intro} and \ref{corrugated},
two time scales control the particle current through the channel:
(i) the bulk relaxation time, $m/\gamma$, and (ii) the ballistic
time across the pore, $m/\gamma_T$. The latter is a measure of
transient effects that may be detected only at the shortest
distances, here, the pore width. To bridge the above time scales
we introduce the effective relaxation time $\tau_{\rm
eff}=m/\gamma_{\rm eff}$, where the effective damping constant is
defined as
\begin{equation}
\gamma_{\rm eff}=\gamma (1+{\gamma_T}/{\gamma}).
\label{th-g}
\end{equation}
Correspondingly, the rescaled mobility function can be approximated to
\begin{equation}
\gamma\mu=\frac{\gamma\mu|_\infty}{1+\gamma_T/\gamma},
\label{th-mob}
\end{equation}
where $\gamma\mu|_\infty$ denotes the rescaled mobility in the
overdamped limit, $\gamma \to \infty$. Despite its being a simple
interpolating formula, Eq.~(\ref{th-mob}) fits quite closely the
simulation curves of Figs.~\ref{F4}(a) and (b) for large drives
(main panel). Note that the horizontal asymptotes for large
$\gamma$ coincide with the expected values of $\gamma\mu|_\infty$,
whose dependence on the compartment geometry, noise and drive
intensity is analytically known
\cite{Bere1,Bere2,PHfest,Borromeo}.

Let us consider next the rescaled mobility at low drives, $F\ll
F_T$. For $F=0$ the transport quantifiers $\gamma \mu_0$ and
$D(0)$ can be formally expressed in terms of the mean exit time,
$\bar \tau_e$, of the Brownian particle out of a single
compartment, namely, $D(0)=x_L^2/4{\bar \tau_e}$ and
$\mu_0=D(0)/kT$, see Eq.~(\ref{einstein}). An analytical
expression for $\bar \tau_e$ as a function of the compartment
geometry is only available in the overdamped dynamics
approximation \cite{Schuss}. In the absence of a fully analytical
treatment, we interpret the numerical results shown in the inset
of Fig.~\ref{F4}(b) by assuming a 1D collisional dynamics along
the $x$-axis. At very low damping and $F=0$, the particle bounces
off the same compartment wall with rate $2v_{\rm th}/x_L$ (attack
frequency), but only a fraction $\Delta/y_L$ of such collisions
leads to a pore crossing. As a consequence, ${\bar \tau_e} \sim
x_Ly_L/2\Delta\sqrt{kT/m}$ and
\begin{equation}
\gamma\mu_0\sim\frac{\gamma x_L}{2\sqrt{mkT}}\frac{\Delta}{y_L} = \frac{r}{2}\frac{\gamma}{\gamma_T},
\label{th-mu0}
\end{equation}
which reproduces with the linear fit in the inset of
Fig.~\ref{F4}(b). Note that such a qualitative argument applies to
the weakly corrugated channels of Fig.~\ref{F2}(c), as well. In
that case, however, $v_{\rm th}$ must be replaced by
$(2/\pi)v_{\rm th}$, to account for an almost isotropic 2D
distribution of the ballistic trajectories inside the compartment;
correspondingly, the factor $1/2$ on the r.h.s. of
Eq.~(\ref{th-mu0}) should be changed to $\pi/4$, see
Fig.~\ref{F2}(d).

The scaling law of the diffusivity at large drives, $D(F) \propto
F^\beta$ with $\beta=2$, can be quantitatively determined by
generalizing an argument originally introduced for the overdamped
regime~\cite{PHfest}. For large $F$, the instantaneous particle
velocity, $ v(t)\equiv \dot x(t)$, switches between a locked mode
with $v_0=0$, as it sticks against a compartment wall, and a
running mode with $v_\infty=F/\gamma$, as it runs along the
central lane of the channel. In view of Eq.~(\ref{mobOO}), it is
clear that the particle spends a fraction $1-\Delta/y_L$ of the
time in the locked mode, and the remaining $\Delta/y_L$ of the
time in the running mode. The random variable $v(t)$ can thus be
modeled as a dichotomic process with subtracted autocorrelation
function~\cite{Gardiner}
\begin{eqnarray}\nonumber
C(t)&\equiv& \lim_{s\to \infty}[\langle v(t+s) \rangle - \langle v
\rangle ][\langle v(s) \rangle - \langle v \rangle ]\\ \nonumber &=&
(v_\infty -v_0)^2 \frac{\bar \tau_0 \bar\tau_\infty}{\bar \tau^2}
~\exp\left(-\frac{\bar{\tau} t}{\bar \tau_0 \bar\tau_\infty}\right),
\end{eqnarray}
where $\bar \tau_0 = (1-\Delta/y_L)\bar \tau$ and $\bar
\tau_\infty = (\Delta/y_L) \bar \tau$  are the average permanence
times, respectively, in the locked and running mode; their sum,
$\bar \tau$, is the relaxation time constant of the dichotomic
process, still to be determined. The spatial diffusivity $D(F)$
can be obtained by integrating  $C(t)$ over time and then making
use of the explicit expressions for $v_0$, $v_\infty$, $\bar
\tau_0$, and $\bar \tau_\infty$, namely
\begin{equation}\label{diffusivity1}
D(F) = \int_0^\infty C(t)dt=\left(\frac{F}{\gamma}\right )^2 \left [\frac{\Delta}{y_L}
\left(1 - \frac{\Delta}{y_L} \right)\right ]^2\bar \tau.
\end{equation}
To determine the unknown time constant $\bar \tau =
(y_L/\Delta) \bar \tau_\infty$, we notice that a
particle remains in the running mode for a time $\bar \tau_\infty$
of the order of the time it takes to diffuse out of the central
channel lane, namely, for low damping,
\begin{equation}\label{tauOO}
2D_0\tau_\infty=\frac{1}{4}[(y_L+\Delta)^2-(y_L-\Delta)^2].
\end{equation}
By inserting the analytical expression for $\tau_\infty$ thus
derived into Eq.~(\ref{diffusivity1}) and taking for simplicity
the limit of narrow pores, $\Delta \ll y_L$, one arrives at
\begin{equation}\label{th-diff}
\frac{D(F)}{D_0}=\frac{\Delta}{2y_L}\left (\frac{F}{F_T}\right )^2.
\end{equation}
This expression is independent of $\gamma$ and well reproduces all
simulation data of Fig.~\ref{F4}(b) at large $F$ or, more
precisely, under the condition that $\gamma \ll \gamma_F$.

On comparing the asymptotic laws for the diffusivity at $\gamma
\to 0$, Eq.~(\ref{th-diff}), and at $\gamma \to \infty$,
Eq.(\ref{th-diffOO}), one would expect $D(\gamma\to 0)/D(\gamma
\to \infty)=\Delta/y_L$. On the contrary, in Fig.~\ref{F4}(c) we
immediately notice that all $D(F)$ curves approach the same
asymptotic scaling law, Eq.(\ref{th-diffOO}). As discussed for the
corrugated channels, the overdamped diffusion scaling law,
Eq.~(\ref{th-diff}), holds only under the condition that $\gamma
\gg \gamma_F$. Correspondingly, here as well, increasing $F$ such
that $F>\gamma^2\Delta/m$, or $l_F \gg \Delta$, makes inertia
effects on confined diffusion emerge (though in a less dramatic
way).

\begin{figure}[tp]
\centering
\includegraphics[width=0.46\textwidth]{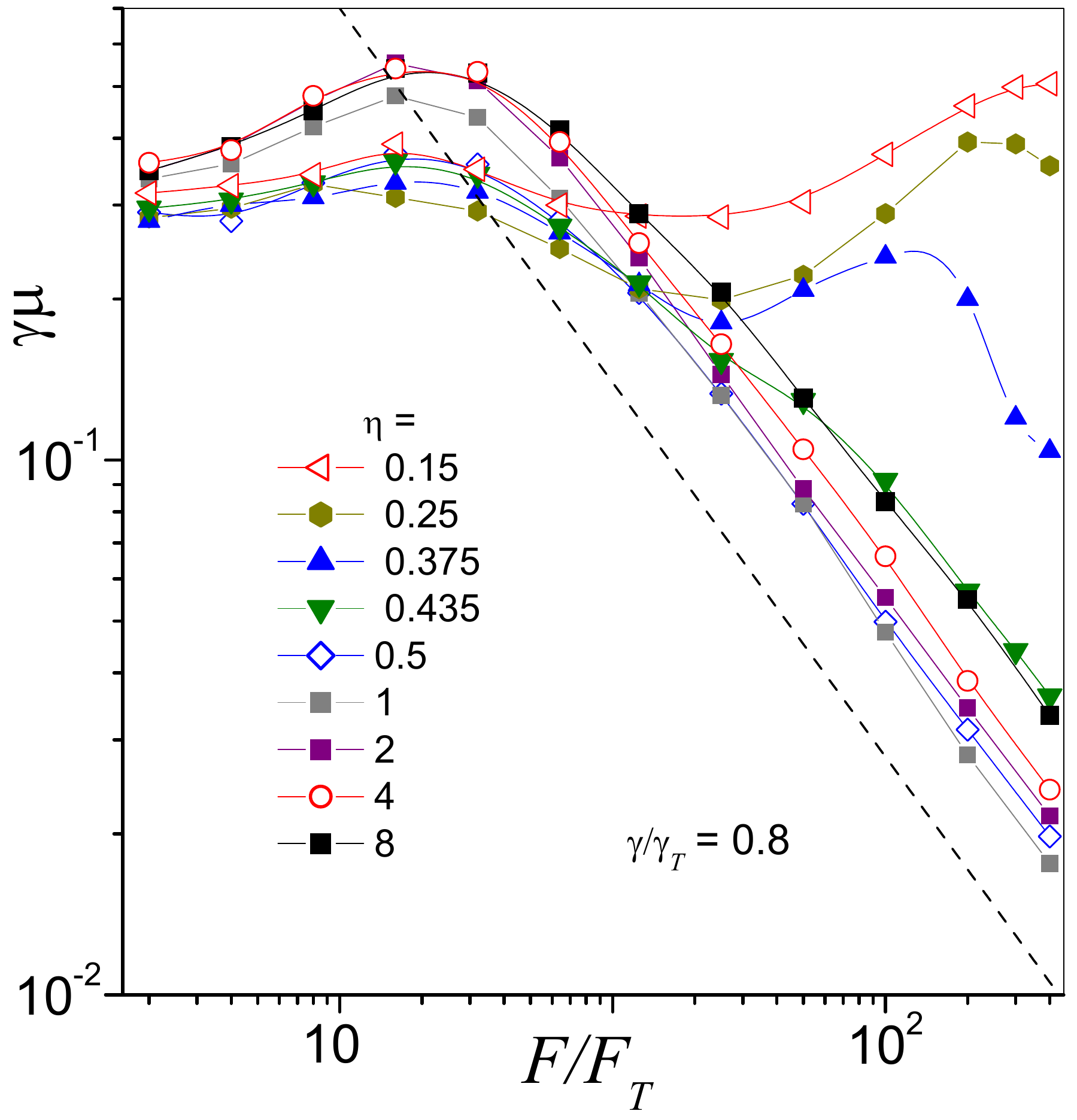}
\caption{(Color online) Channel with tunable corrugation,
Eq.~(\ref{w-eta}): rescaled mobility, $\gamma \mu$ versus  $F/F_T$
for different $\eta$. Other simulation parameters: $r=1$,
$\Delta/y_L=0.1$, and $\gamma/\gamma_T=0.8$. The dashed line is
the power-law $(F/F_T)^{-\alpha/2}$, with $\alpha=1.4$ drawn in
Fig.~\ref{F2}(b) for $\eta=2$. \label{F5}}
\end{figure}

\section{Discussion} \label{discussion}

The comparison between transport properties in smoothly and
sharply corrugated channels is suggestive of the role played by
the channels profile in the presence of inertia. In principle,
both channel geometries of Sec.~\ref{corrugated} and \ref{septate}
could be reproduced by means of one parametric profile function,
say,
\begin{equation}
\label{w-eta}
w_\eta(x)=\frac{1}{2}\left [\Delta+(y_L-\Delta)\sin^{\eta}\left (\frac{\pi x}{x_L}\right) \right ],
\end{equation}
with tunable exponent $\eta$ \cite{Ghosh note}. This function
coincides with $w(x)$, Eq.~(\ref{walls}), for $\eta=2$ and
approaches a rectangular compartment $x_L\times y_L$ for $\eta \to
0$. The divide between smoothly and sharply corrugated channels
can be set at $\eta=1$, where the two sides of the bottleneck
profile change from concave, for $\eta>1$, to convex, for
$\eta<1$. Such change in the pore geometry affects, for instance,
the drive dependence of the rescaled mobility at {\it low}
damping, see Fig.~\ref{F5}. All curves $\gamma \mu(F)$ with
$\eta>1$ decay with the same approximate power law as reported in
Fig.~\ref{F2}(b) for $\eta=2$. For $\eta<1$, instead, the behavior
of $\gamma \mu(F)$ is as in Fig.~\ref{F2}(b) (sinusoidally
corrugated channel), at low $F$, and in Fig.~\ref{F4}(a) (septate
channel), at large $F$. Without further analyzing the $\eta$
dependence of the transport quantifiers, we now discuss certain
differences and similarities between sinusoidal and septate
channels.

In both types of channels, the diffusivity grows asymptotically
with the drive according to a power law, $D(F) \propto F^\beta$,
where $\beta=2$ for septate channels and $\beta=1$ for sinusoidal
channels. For smoothly corrugated channels this result may come as
a surprise, since, for large damping, the diffusivity is known to
approach its bulk value, $D(F\to \infty)=D(0)$. The asymptotic
power law with $\beta=1$ illustrated in the inset of Fig.~\ref{F3}
results from the enhanced trapping effect due to the underdamped
particle bouncing back and forth against the compartment walls. A
simple qualitative argument yields~\cite{our EPL}
\begin{equation}\label{corrugated-diff}
\frac{D(F)}{D_T} \sim \frac{\pi}{8}\frac{x_L F}{kT}=r\frac{\pi}{8}\frac{y_L}{\Delta}\frac{F}{F_T},
\end{equation}
in rather good agreement with our simulation data (see inset of
Fig.~\ref{F3}). Note that, lowering the temperature, for small
damping $D(F)$ diverges like $T^{-1/2}$, which means that
diffusion is the result of chaotic ballistic collisions.

Inertia corrections to the drive dependence of the diffusivity are
not as dramatic in septate channels, Fig.~\ref{F4}(c), as they
appear in corrugated channels, Fig.~\ref{F3}. This explains why
the role of the threshold $\gamma_F$, Eq.~(\ref{gF}), is less
prominent for sharply corrugated channels. On a closer look,
however, one notices that, on increasing $\gamma$, the data points
for $D(F)/D_0$ approach the predicted oblique asymptote of
Eq.~(\ref{th-diff}) at larger and larger $F$, consistently with
the large-drive inertial regime $\gamma \ll \gamma_F$.

Finally, we notice that septate and sinusoidally corrugated
channels also differ in the large-drive behavior of their
mobilities at low damping. While in septate channels $\gamma
\mu(F\to \infty)$ was shown to approach a small but finite value,
$\gamma \mu|_\infty (\gamma/\gamma_T)$, see Eq.~(\ref{th-mob}),
the mobility in sinusoidal channels was numerically fitted by the
scaling law $\gamma \mu(F\to \infty) \sim
(\gamma/\gamma_F)^\alpha$, where $\alpha$ is an increasing
function of $\Delta$ with $\alpha(\Delta\to 0)=2$
[Fig.~\ref{F2}(c)]. This means that in sinusoidal channels
$\langle v(F)\rangle$ [and not $\gamma \mu(F)$] tends to a finite
asymptotic value. By definition, the rescaled mobility can be
written as $\gamma \mu \sim (x_L/\bar \tau_d)(\gamma/F)$, where
$\bar \tau_d$ denotes the mean drift time of a particle across a
compartment in the presence of a strong drive. Accordingly, as
$\alpha \to 2$, the drift time $\bar \tau_d$ becomes insensitive
to the (large) drive, which hints at an emerging ballistic
dynamics \cite{gaspard}.
We also remark that the above scaling law for $\gamma \mu(F\to
\infty)$ applies to all pore geometries with $\eta \geq 1$
\cite{Ghosh note}, see Fig.~\ref{F5}; for septate channels,
 $\eta \to 0$, such a scaling law, with $1 < \alpha < 2$, closely reproduces the decaying branch of the mobility curves displayed in Fig.~\ref{F4}(a).

\section{Conclusions} \label{conclusion}

The main result of this work is that for real physical suspensions
flowing through confined geometries, both in biological and
artificial systems, pore crossings grow increasingly sensitive to
the suspension fluid viscosity with decreasing the pore radius.
With respect to previous attempts at incorporating finite-mass
effects in the analysis of Brownian transport through corrugated
narrow channels~\cite{inertia1,inertia2,inertia3}, we stress that
the inertial effects reported here are not of mere academic
interest~\cite{our EPL}.

Inertial effects can be directly observed, for instance, in a
dilute solution of colloidal particles driven across a porous
membrane or an artificial sieve~\cite{Mark,Garbow}. On the other
hand, channel profiles at the micro- and nanoscales can be
tailored as most convenient~\cite{Dekker}. As detailed in
Ref.~\cite{our EPL}, the experimental  demonstration of inertial
effects on Brownian  transport through narrow pores is to become
accessible when manipulating artificial particles of micrometric
size by means of well established experimental
techniques~\cite{Li,Huang,Franosch,Jannasch}. For nano-particles,
like biological molecules, detecting such effects will require
more refined experimental setups.

\section*{Acknowledgements}

This work was partly supported  by the European Commission under
grant No. 256959 (NANOPOWER) (FM), the Volkswagen foundation
project I/83902 (PH, GS), the German excellence cluster
"Nanosystems Initiative Munich'' (NIM) (PH, GS),  the Augsburg
center for Innovative Technology (ACIT) of the University of
Augsburg (FM, PH), and the Japanese Society for Promotion of
Science (JSPS) through Fellowship No. P11502 (PKG) and No. S11031
(FM). FN is partially supported by the ARO, NSF grant No. 0726909,
JSPS-RFBR contract No. 12-02-92100, Grant-in-Aid for Scientific
Research (S), MEXT Kakenhi on Quantum Cybernetics, and the JSPS
via its FIRST program.

\end{document}